\documentclass[twocolumn,prl,aps,epsf,showpacs]{revtex4}

\usepackage{graphicx}
\usepackage{dcolumn}
\usepackage{bm}

\begin{document}

\title{Wigner Crystallization in a Quasi-3D Electronic System}

\author{B.A. Piot$^{1}$, Z. Jiang$^{2,3}$, C.R. Dean$^{1}$, L.W. Engel$^{2}$, G. Gervais$^{1}$, L.N. Pfeiffer$^{4}$ and K.W. West$^{4}$ }

\affiliation{$^{1}$Department of Physics, McGill University,
Montreal, H3A 2T8, CANADA}

\affiliation{$^{2}$National High Magnetic Field Laboratory,
Tallahassee, Florida 32310, USA}

\affiliation{$^{3}$Department of Physics, Columbia University, New York, New York 10027,
USA}

\affiliation{$^{4}$Bell Labs, Alcatel-Lucent Incorporation, Murray
Hill, New Jersey 07974, USA}

\affiliation{\vspace{0.5cm} \textsf{Nature Physics 4, 936 - 939
(2008) Published online: 5 October 2008
http://www.nature.com/nphys/journal/v4/n12/abs/nphys1094.html}}

 \maketitle

\textbf{When a strong magnetic field is applied perpendicularly
(along z) to a sheet confining electrons to two dimensions (x-y),
highly correlated states emerge as a result of the interplay
between electron-electron interactions, confinement and disorder.
These so-called fractional quantum Hall (FQH) liquids
\cite{TsuiFQH82} form a series of states which ultimately give way
to a periodic electron solid that crystallizes at high magnetic
fields. This quantum phase of electrons has been identified
previously as a disorder-pinned  two-dimensional Wigner crystal
with broken translational symmetry in the x-y plane
\cite{WignerPhysRev.46.1002,Lozoviktheory75,Lamtheory84,Levesquetheory84,WilliamsGlatti91PRL.66.3285,electronsolidreview,ChenMeltingsolid06}.
Here, we report  our discovery of a new insulating quantum phase
of electrons when a very high magnetic field, up to 45T, is
applied in a geometry \emph{parallel} (y-direction) to the
two-dimensional electron sheet. Our data point towards this new
quantum phase being an electron solid in a `quasi-3D'
configuration induced by orbital coupling with the parallel
field.}

The formation of an electron solid has been observed previously at
very high magnetic fields where less than $\frac{1}{5}$ of the
lowest Landau level describing the orbital dynamics is occupied by
electrons, or at zero magnetic field in extremely dilute 2D
systems realized on helium surfaces
\cite{GrimesclassicalsolidPRL.42.795}. Recently,
a 1D Wigner crystal  was also reported for electrons
in carbon nanotubes
\cite{Bockrath.2008}. Here, we present evidences for another
possibility where the crystallization would occur in a `quasi-3D'
electronic system, evolving continuously from a disorder-pinned 2D
state.  In our work, a  two-dimensional electron gas  (2DEG) is
rotated inside a magnetic field by a tilting angle $\theta$, so that an
in-plane magnetic field $B_{\parallel}$ is added parallel to
the 2DEG. Provided this field be large enough, it has been
proposed theoretically \cite{Yu02} that the  electron solid energy
would become lower than that of the FQH liquid at Landau level
filling factors $\nu=n_{s}/(eB_{\perp}/h)$, where $n_{s}$ is the electron density,
significantly higher than for the conventional 2D Wigner crystal in which $\nu\lesssim
\frac{1}{5}$. A tilt-induced liquid-to-insulator transition has
been observed in a 2D hole system \cite{pan05tiltholesolid, Santos92holeone3,csathyholephasediagram04} in
which the insulating phase, initially located at $\nu \lesssim
\frac{1}{3}$, was shifted close to $\nu=\frac{2}{3}$ with
increasing $\theta$. However, such a
tilt-induced transition for electrons has so far been elusive.
Here, combining a  relatively `thick' high-mobility  2DEG and very
high parallel magnetic fields $B_{\parallel}$ to enhance
orbital coupling, we report a transition from a FQH liquid to an
insulating phase in an \emph{electron} system.

\begin{figure}[tbp]
\includegraphics[width=0.8\linewidth,angle=0,clip]{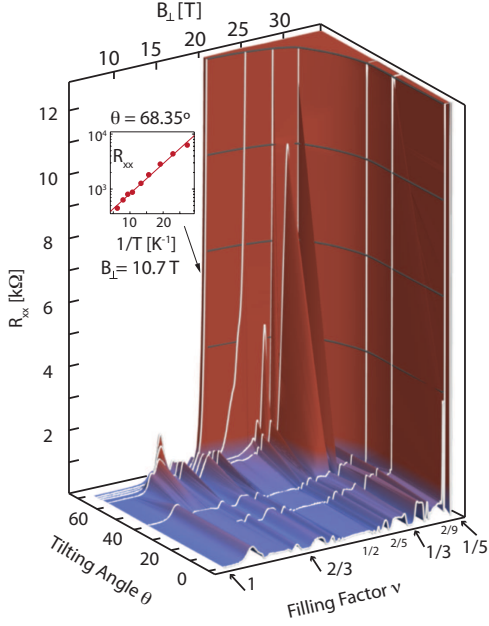}
\caption{Longitudinal resistance $R_{xx}$ versus tilt and filing factor $\nu$. The resistance
of sample C is shown as a function of the
perpendicular magnetic field $B_{\perp}$ (or equivalently the  filling
factor $\nu$) for different tilting angles $\theta$ and  T $\simeq$
35 mK. White lines are actual data, and the trend in between is
extrapolated as a guide-to-the-eye. The resistance value is
emphasized by a colourmap, with the black lines corresponding to
equi-resistance. Inset: temperature dependence of $R_{xx}$ at
$\theta$=68.35$^\circ$ and $B_{\perp}$=10.7 T. }\label{fig1}
\end{figure}

In Fig.~\ref{fig1}, we report in a 3D plot the longitudinal
resistance, $R_{xx}$ of sample C as a function of the
perpendicular magnetic field for different tilting angles. In the
perpendicular configuration, {\it i.e.} when $\theta$=0$^{\circ}$,
we observe the usual  FQH series characterized by a vanishing
longitudinal resistance due to the opening of a many-body gap in
the density of states. At higher perpendicular magnetic fields,
for filling factors lower than $\nu=\frac{1}{5}$, a divergence of
$R_{xx}$ corresponding to the onset of a deep insulating state
($R_{xx} \gtrsim 800 k\Omega$), previously identified as a
disorder-pinned Wigner Crystal, is observed
\cite{WilliamsGlatti91PRL.66.3285,electronsolidreview,ChenMeltingsolid06}.
A reentrance of the insulating state ($R_{xx} \sim 400 k \Omega$)
is also observed in a narrow region between $\nu=\frac{1}{5}$ and
$\nu=\frac{2}{9}$ \cite{Jiang90}. As the tilting angle is further
increased, the onset of the divergence shifts to lower
perpendicular fields, signaling the appearance of an insulating
phase at higher filling factors, and approaching $\nu=\frac{2}{3}$
at $\theta=68.35^{\circ}$. The temperature dependence of $R_{xx}$
at $\theta=68.35^{\circ}$ and $B_{\perp}$=10.7 T is shown in the
inset. At this magnetic field, $R_{xx}$ increases exponentially
with $1/T$, as observed in the zero tilt insulating phase below
$\nu=\frac{1}{5}$ \cite{Willett88,Jiang90}, thus confirming the
similarity of this new quantum phase with  the 2D Wigner crystal
phase at zero tilt. At this angle, we also observed a critical
field $B_{\perp, c}$ separating a metallic ($\partial R_{xx} /
\partial T >0 $) from an insulating region ($ \partial
R_{xx}/\partial T <0 $) for which $R_{xx}$ does not depend on
temperature, characteristic of a metal-to-insulator
transition\cite{Shahar}. We also observe FQH states (such as at
$\nu=1/3$) that remain robust for intermediate angles even though
the neighbouring (higher) filling factor region becomes
insulating, {\it  i.e.} showing up as resistance peaks in
Fig.~\ref{fig1}. This is reminiscent of the re-entrant insulating
phase observed in the pure perpendicular field case.

\begin{figure}[tbp]
\includegraphics[width=0.8\linewidth,angle=0,clip]{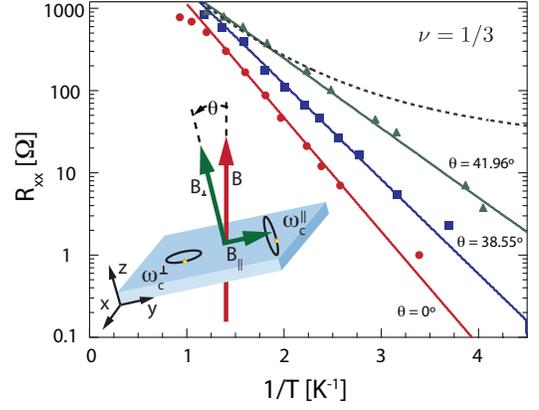}
\caption{Thermal activation plots at $\nu=1/3$ for sample A. The
longitudinal resistance $R_{xx}$ is plotted as a function of $1/T$
for different tilting angles $\theta$ (solid symbols, error bars smaller
than symbols). The solid
lines are fit enabling the extraction of the thermally activated FQH
gaps. Calculation of $R_{xx}$ for (pure) disorder-induced  gap suppression
 at 41.96$^{\circ}$ is shown with a dashed line (see text).
}\label{fig2}
\end{figure}

\begin{figure}[tbp]
\includegraphics[width=0.86\linewidth,angle=0,clip]{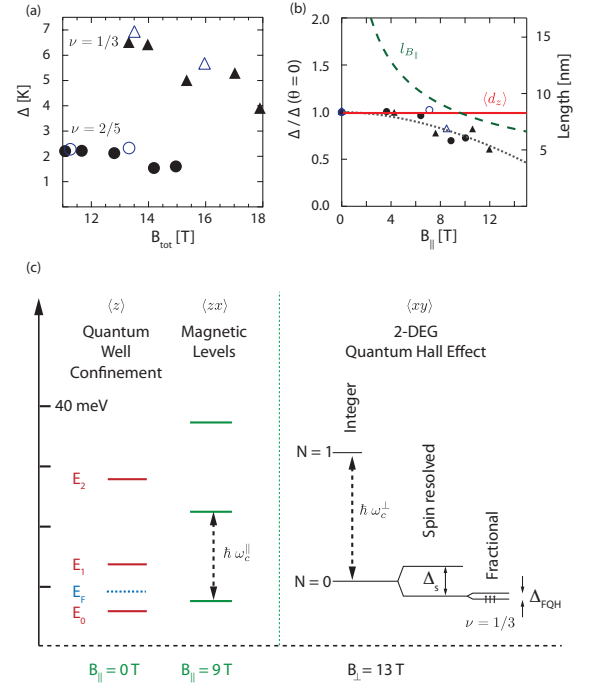}
\caption{Activation energy gaps and relevant  energy scales. (a) Thermally activated gap at $\nu=1/3$ (triangles) and
$\nu=2/5$ (circles) as a function of the total magnetic field
$B_{total}$. Sample A (closed symbols) and B (open symbols). (b)
Thermally activated gap normalized with respect to their
$\theta=0^{\circ}$ value, at $\nu=1/3$ (triangles) and $\nu=2/5$
(circles) as a function of the parallel magnetic field
$B_{\parallel}$. The dotted line is a guide to the eye. Magnetic
length $l_{B_{\parallel}}$ associated with $B_{\parallel}$ (dashed
line) and average thickness of the 2DEG in the \textbf{z}
direction $<d_{z}>$ (solid line) (right scale). All data error bars smaller
than symbols. (c) Energy diagram
(see text). All energies are calculated with respect to the bottom
of the conduction band (dashed horizontal line) and are shown to
scale.}\label{fig3}
\end{figure}

To further investigate the nature of this new insulating quantum
phase, we have measured the FQH gap at
$\nu=\frac{1}{3},\frac{2}{5}$ as a function of the tilting angle
$\theta$. In Fig.~\ref{fig2} the longitudinal resistance $R_{xx}$
of sample A at $\nu=\frac{1}{3}$ is plotted on a semi-log scale as
a function of the inverse temperature $1/T$, for various tilting
angles. $R_{xx}$ clearly follows a thermally activated behaviour
where $R_{xx} \propto e^{\frac{-\Delta}{2k_{B}T}}$ over a few
decades, allowing us to extract the thermally activated gap,
$\Delta$. The resulting values are reported in Fig.~\ref{fig3}a as
a function of the total magnetic field, $B_{total}$. At high
tilting angles (total magnetic field), both the $\frac{1}{3}$ and
$\frac{2}{5}$ gaps are reduced. For the $\frac{1}{3}$ FQH state,
the weakening of the gaps {\it cannot} be accounted for  by spin
effects since its ground state is fully spin-polarized. Similarly,
at the large $B_{\perp}$ involved here, the $\nu= \frac{2}{5}$ FQH
state is most likely spin-polarised
\cite{LeadleyCompskyrmion97,Kang97}, so no spin-related effects
are expected to reduce the gap. In addition, pure disorder-related
effects occurring at high $\theta$ \cite{DasSarmaHwang2000} cannot
account for the gap reduction observed here. The calculation of
$R_{xx}$ at $41.96^{\circ}$ with disorder-induced gap suppression
owing to an increase in Landau level broadening \cite{Usher91}
(dashed line in Fig.~\ref{fig2}) does not qualitatively agree with
the observed linear trend. The Landau level broadening used to fit
the onset of the $R_{xx}$ decrease is seven times larger than
measured from the Shubnikov de Haas oscillations at zero tilt.
This is inconsistent with the observation of well-developed FQH
states at this angle, $\theta=41.96^{\circ}$.

In Fig.~\ref{fig3}b we plot the same thermally activated gap
normalized by their value at $\theta=0^{\circ}$, as a function of
the parallel magnetic field $B_{\parallel}$. Interestingly, a
similar collapse of the FQH states is obtained as a function of
$B_{\parallel}$. More strikingly, the gap reduction begins at
$B_{\parallel}\sim$ 7T which is very close to the parallel
magnetic field $B_{\parallel}\sim$ 9T where the magnetic length
$l_{B_{\parallel}}=\sqrt{\hbar/eB_{\parallel}}=8.5$ nm associated
with $B_{\parallel}$ approaches the average thickness of the 2DEG
in the \textbf{z} direction, $<d_{z}>=8.3$ nm (calculated using
Ref.\cite{StopasquareQW92}).
This is strong evidence for the collapse of the FQH
 gaps to be directly related to orbital coupling with the parallel
field, becoming stronger when electrons are confined to
dimensions smaller than the initial 2DEG thickness.
For  $B_{\parallel}\gtrsim 10$T, the electron gas now defines a  `quasi-3D' system, with the quantum
well now primarily providing a sample with a finite width in the \textbf{z}
direction.

One can obtain an estimate for the angle $\theta_{c_{1/3}}$ for
which the $\nu=\frac{1}{3}$ gap vanishes completely by
extrapolating the $B_{\parallel}$ trend observed in
Fig.~\ref{fig3}b. The resulting angle, $\theta_{c_{1/3}}^{gap}\sim
58\pm 5^{\circ}$, correlates very well with the angle at which the
insulating phase appears at $\nu=\frac{1}{3}$, estimated from
Fig.~\ref{fig1} to be $\theta_{c_{1/3}}^{Rxx} \sim61. \pm 2
^{\circ}$. This reinforces our interpretation for the collapse of
the FQH gap at $\nu=\frac{1}{3}$, and the related emergence of an
insulating phase to be driven by orbital coupling to the parallel
field.

It is instructive to consider all relevant energies as tilt is
increased, and these are depicted schematically in
Fig.~\ref{fig3}c. In the \textbf{x-y} plane, the conventional
quantum Hall effects and associated Landau levels are separated by
the cyclotron energy $\hbar\omega_{c}^{\perp}=\hbar
eB_{\perp}/m^{*}$, represented here for the perpendicular magnetic
field where the $\nu=\frac{1}{3}$ FQH state is observed in sample
A, $B_{\perp} \sim$ 13 T. The spin gap, $\Delta_{s}$,  is shown
for the $N=0$ Landau level, and $\Delta_{FQH}$ gives the magnitude
of the $\nu=\frac{1}{3}$ FQH gap.  Along the confinement axis
\textbf{z}, we use calculations for the confinement energy of a 40
nm wide quantum well as given in Ref.\cite{Stern85}, and report
here the first three electric subbands, $E_{0}$, $E_{1}$ and
$E_{2}$. The Fermi energy $E_{F}$ is indicated in absence of a
magnetic field, showing that all electrons are confined in the
first subband, $E_{0}$. When a sufficient parallel magnetic field
is applied, magnetic levels take over the confinement levels in
determining the \textbf{z} energy;  these  `Landau-like'  levels
are separated by the `parallel cyclotron energy'
$\hbar\omega_{c}^{\parallel}$ and they determine the energy in the
\textbf{z-x} plane \cite{StopaBparaLandaulevelPRB.40.10048}. When
the parallel magnetic field is further increased, the occupation
of the lowest `Landau-like' magnetic level with a degeneracy
$eB_{\parallel}/h$ is reduced. We note, however, that we
described for simplicity two independent sub-systems in the
\textbf{x-y} and \textbf{z-x} planes, whereas in reality they are
naturally coupled in the \textbf{x} direction.

\begin{figure}[tbp]
\includegraphics[width=0.86\linewidth,angle=0,clip]{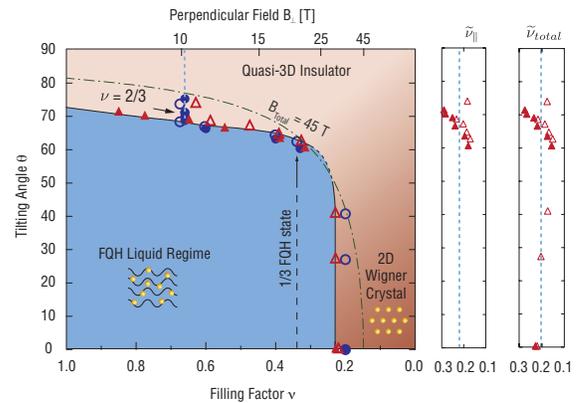}
\caption{Phase diagram for the `quasi-3D' insulator at $T\simeq
35$mK. Critical filling factors $\nu_{c_{1}}$ (circles) and
$\nu_{c_{2}}$ (triangles) (defined in text) as a function of the
tilting angle $\theta$. The open symbols are data extracted from
Fig.~\ref{fig1}, and the closed symbols were obtained on a separate
cooldown. The $B_{total}$=45 T line (dash-dot) determines the
experimental observable range. Right panel:   `parallel (total)
filling factor' $\tilde{\nu}_{\parallel}$ ($\tilde{\nu}_{total}$)
(defined in the text) associated with the critical filling factor
$\nu_{c_{2}}$ (triangles). The vertical dashed line is the average
value of $\tilde{\nu}_{\parallel}$ ($\tilde{\nu}_{total}$).}
\label{fig4}
\end{figure}

From the data in Fig.~\ref{fig1}, we can construct a phase diagram
for the `quasi-3D'  insulator at high tilt (parallel field).
For this,  we (arbitrarily) define a critical filling factor
$\nu_{c_{1}}$ corresponding to the smallest $\nu$ value for which
a FQH state is observed, as well as a critical filling factor
$\nu_{c_{2}}$ corresponding to the largest $\nu$ value for which
the resistance value exceeds $h/2e^2$=12.91 k$\Omega$. One can therefore view
$\nu_{c_{1}}$ as the liquid phase termination, and $\nu_{c_{2}}$
as the onset of the insulating phase.  The criterion for $\nu_{c_{2}}$
is justified by the inset of Fig.\ref{fig1} which
shows the sample to be already  insulating for this resistance value. In this context, the
reentrance of the insulating phase, usually observed between the
$\nu=\frac{2}{9}$ and $\nu=\frac{1}{5}$, is characterized by
$\nu_{c_{2}} > \nu_{c_{1}}$. These critical filling factors
$\nu_{c_{1}}$ (circles)  and $\nu_{c_{2}}$ (triangles)  are plotted in Fig.~\ref{fig4} as
a function of the tilting angle $\theta$, and at base temperature
$T\simeq35$ mK, with closed and open symbols denoting data obtained
during two separate cooldowns. As the tilt angle $\theta$ is
increased, a higher total magnetic field is required to achieve
the perpendicular field necessary to observe the $\nu=\frac{1}{5}$
FQH state and the neighbouring insulating region. This restricts
our tilting  range for that state to angles $\theta \lesssim
42^{\circ}$, within the 45T of our magnet indicated as a
dash-dotted line. However, further tilt of the sample shifts the
insulating phase to higher filling factors, so that it reappears
within our field range at angles $\theta \gtrsim 60^{\circ}$. The
persistence of the $\nu=\frac{2}{3}$ FQH state at $\theta \gtrsim
70 ^{\circ}$ results in having a reentrant phase
($\nu_{c_{2}}>\nu_{c_{1}}$) in this region of the phase diagram.

The smooth crossover between the low-$\theta$ and high-$\theta$
regions of the phase diagram is
evidence for a  continuous distortion of the
perpendicular-field dominated 2D Wigner crystal  to  an
 electron solid stabilized in a  `quasi-3D' geometry.
The steep divergence of $R_{xx}$ preceding the `quasi-3D'
insulator, as well as the existence of reentrant states, are
analogous to what is observed for the 2D Wigner crystal in the
perpendicular configuration. Recent
numerical calculations \cite{Yu02} have predicted the ground state energy of the
solid phase to be very close to the liquid under sufficient
tilting angles, {\it e.g.}  $\theta\gtrsim 45^{\circ}$ at $\nu=\frac{1}{3}$. For strong parallel fields,
Landau-like magnetic levels in the {\bf z-x}  plane should be
forming as the 2DEG becomes a `quasi-3D' system. The non-zero
perpendicular field $B_{\perp}$ is nevertheless absolutely
required  here to prevent the electronic system from being free along
the \textbf{y} direction and the Landau quantization to be smeared
out by the free kinetic energy. At high tilt angles,  $B_{\perp}$
thus provides an effective magnetic confinement in the
\textbf{y}-direction  so as to recreate a pseudo-2D system along
\textbf{z-x}. In this regime, $\theta\gtrsim 60^{\circ}$, we
can define  a  {\it parallel} filling factor associated with the parallel
field component,
$\tilde{\nu}_{\parallel}=n_{s}/(eB_{\parallel}/h)$ describing
the occupation in the first {\bf z-x} magnetic level. Here, we have
assumed $n_{s}$ to remain unmodified by the field axis. The right
panel of Fig.~\ref{fig4} shows a scatter plot for the values
obtained for $\tilde{\nu}_{\parallel}$ at the onset of the
insulating phase $\nu_{c_{2}}$ and  $\theta \gtrsim 60^{\circ}$,
where $l_{B_{\parallel}}\lesssim 6$ nm.
The transitions from a FQH liquid to an insulating
phase in this regime occur at an average
parallel filling factor $<\tilde{\nu}_{\parallel}>=0.22$, shown by
a dashed vertical line.  We also define a  {\it total}  filling factor
$\tilde{\nu}_{total}\equiv n_{s}/(eB_{total}/h)$,  equal to
$\tilde{\nu}_{\parallel}$ ($\nu$) in the high $B_{\parallel}$
($B_{\perp}$) limit. Using this definition,
 the transitions observed here  occur close to $\frac{1}{5}$ over the
whole $\theta$ range, $\theta=[0,70^{\circ}]$ (right panel in
Fig.~\ref{fig4}). The values of
$\tilde{\nu}_{total}$ at the transitions are similar to those
associated with the onset of the conventional 2D \textbf{x-y}
Wigner crystal, at $\nu< \frac{1}{5}$, therefore suggesting a possible stabilization of an electron solid by the {\it total} magnetic field due to a reduction of the occupation of the quantized orbital energy level. A continuous evolution of the liquid-insulator transitions can be seen in the main panel of Fig.~\ref{fig4}, where the phase boundary clearly mimics that of the total field. The solid phase would occur for  $l_{B_{\parallel}}$ sufficiently small relative to the quantum well width,
so that  electrons may acquire the freedom to minimize their mutual repulsion by adjusting their
positions in the {\bf z}-direction.
We note, however, that the exact structure of this electron solid cannot at present be conjectured and remains an open question which will be better addressed by microwave experiments probing the pinning modes of the crystal.

This work has been supported by the Natural Sciences and
Engineering Research Council of Canada (NSERC), the Canada Fund
for Innovation (CFI), the Canadian Institute for Advanced Research
(CIFAR), FQRNT (Qu\'ebec), the Alfred P. Sloan Foundation (G.G.), and
the NSF under DMR-03-52738 (Z.J.).  We thank
H.L. Stormer and D.C. Tsui for helpful discussions, and J. Hedberg,
G. Jones, T. Murphy and E. Palm for
technical assistance. A portion of this work was performed at the
National High Magnetic Field Laboratory, which is supported by NSF
Cooperative Agreement No. DMR-0084173, by the State of Florida,
and by the DOE.\\

\begin{small}

\end{small}

\vspace*{10mm}

{\bf Methods}

The 2DEGs studied here are  40 nm wide
modulation-doped GaAs quantum wells, all from the same wafer grown by molecular beam epitaxy.
They are referred to as
samples A, B and C, and have densities $n_{s}$=1.05, 1.06 and
1.52$\times10^{11}~\text{cm}^{-2}$ and corresponding mobilities $\mu$= 12(2), 8(2) and
14(2) $\times10^{6}~\text{cm}^{2}/\text{V$\cdot$s}$, respectively. The
samples were cooled in a dilution fridge with base temperature $T\simeq
35$ mK installed inside a hybrid superconducting/resistive magnet capable of reaching a total
magnetic field of 45T. Treatment with a red LED was used during
the cooldown. Transport measurements were performed using a
standard low-frequency lock-in technique at low excitation current, $I_{exc}\sim 2-100$ nA.
The sample was tilted with respect to the total magnetic field $B_{total}$ (see lower
left cartoon in Fig.~\ref{fig2}) using an in-situ rotation stage.
The tilting angle $\theta$ was determined from the shift of the
resistance minimum of well-known integer quantum Hall states,
according to $B_{\perp}=B_{total}\cdot cos(\theta)$.


\begin{thebibliography}{21}
\expandafter\ifx\csname
natexlab\endcsname\relax\def\natexlab#1{#1}\fi
\expandafter\ifx\csname bibnamefont\endcsname\relax
  \def\bibnamefont#1{#1}\fi
\expandafter\ifx\csname bibfnamefont\endcsname\relax
  \def\bibfnamefont#1{#1}\fi
\expandafter\ifx\csname citenamefont\endcsname\relax
  \def\citenamefont#1{#1}\fi
\expandafter\ifx\csname url\endcsname\relax
  \def\url#1{\texttt{#1}}\fi
\expandafter\ifx\csname
urlprefix\endcsname\relax\def\urlprefix{URL }\fi
\providecommand{\bibinfo}[2]{#2}
\providecommand{\eprint}[2][]{\url{#2}}

\bibitem[{\citenamefont{Tsui et~al.}(1982)\citenamefont{Tsui, Stormer, and
  Gossard}}]{TsuiFQH82}
\bibinfo{author}{\bibfnamefont{D.~C.} \bibnamefont{Tsui}},
  \bibinfo{author}{\bibfnamefont{H.~L.} \bibnamefont{Stormer}},
  \bibnamefont{and} \bibinfo{author}{\bibfnamefont{A.~C.}
  \bibnamefont{Gossard}},
   \textit{\bibinfo{bla bla bla}{Two-dimensional magnetotransport in the extreme quantum limit}},
   \bibinfo{journal}{Phys. Rev. Lett.}
  \textbf{\bibinfo{volume}{48}}, \bibinfo{pages}{1559-1562} (\bibinfo{year}{1982}).

\bibitem[{\citenamefont{Wigner}(1934)}]{WignerPhysRev.46.1002}
\bibinfo{author}{\bibfnamefont{E.}~\bibnamefont{Wigner}},
   \textit{\bibinfo{bla bla bla}{On the interaction of electrons in metals}},
  \bibinfo{journal}{Phys. Rev.} \textbf{\bibinfo{volume}{46}},
  \bibinfo{pages}{1002-1011} (\bibinfo{year}{1934}).

\bibitem[{\citenamefont{Lozovik and Yudson}(1975)}]{Lozoviktheory75}
\bibinfo{author}{\bibfnamefont{Y.~E.} \bibnamefont{Lozovik}} \bibnamefont{and}
  \bibinfo{author}{\bibfnamefont{V.~I.} \bibnamefont{Yudson}},
     \textit{\bibinfo{bla bla bla}{Crystallization of a two-dimensional electron gas in a magnetic field }},
  \bibinfo{journal}{JETP Lett.} \textbf{\bibinfo{volume}{22}},
  \bibinfo{pages}{11-12} (\bibinfo{year}{1975}).

\bibitem[{\citenamefont{Lam and Girvin}(1984)}]{Lamtheory84}
\bibinfo{author}{\bibfnamefont{P.~K.} \bibnamefont{Lam}} \bibnamefont{and}
  \bibinfo{author}{\bibfnamefont{S.~M.} \bibnamefont{Girvin}},
     \textit{\bibinfo{bla bla bla}{Liquid-solid transition and the fractional quantum-Hall effect}},
  \bibinfo{journal}{Phys. Rev. B} \textbf{\bibinfo{volume}{30}},
  \bibinfo{pages}{473-475} (\bibinfo{year}{1984}).

\bibitem[{\citenamefont{Levesque et~al.}(1984)\citenamefont{Levesque, Weis, and
  MacDonald}}]{Levesquetheory84}
\bibinfo{author}{\bibfnamefont{D.}~\bibnamefont{Levesque}},
  \bibinfo{author}{\bibfnamefont{J.~J.} \bibnamefont{Weis}}, \bibnamefont{and}
  \bibinfo{author}{\bibfnamefont{A.~H.} \bibnamefont{MacDonald}},
     \textit{\bibinfo{bla bla bla}{Crystallization of the incompressible quantum-fluid state of a two-dimensional electron gas in a strong magnetic field}},
  \bibinfo{journal}{Phys. Rev. B} \textbf{\bibinfo{volume}{30}},
  \bibinfo{pages}{1056-1058} (\bibinfo{year}{1984}).

\bibitem[{\citenamefont{Williams et~al.}(1991)\citenamefont{Williams, Wright,
  Clark, Andrei, Deville, Glattli, Probst, Etienne, Dorin, Foxon
  et~al.}}]{WilliamsGlatti91PRL.66.3285}
\bibinfo{author}{\bibfnamefont{F.~I.~B.} \bibnamefont{Williams}},
  \bibinfo{author}{\bibfnamefont{P.~A.} \bibnamefont{Wright}},
  \bibinfo{author}{\bibfnamefont{R.~G.} \bibnamefont{Clark}},
  \bibinfo{author}{\bibfnamefont{E.~Y.} \bibnamefont{Andrei}},
  \bibinfo{author}{\bibfnamefont{G.}~\bibnamefont{Deville}},
  \bibinfo{author}{\bibfnamefont{D.~C.} \bibnamefont{Glattli}},
  \bibinfo{author}{\bibfnamefont{O.}~\bibnamefont{Probst}},
  \bibinfo{author}{\bibfnamefont{B.}~\bibnamefont{Etienne}},
  \bibinfo{author}{\bibfnamefont{C.}~\bibnamefont{Dorin}},
  \bibinfo{author}{\bibfnamefont{C.~T.} \bibnamefont{Foxon}}, \bibnamefont{and}
    \bibinfo{author}{\bibfnamefont{J.~J.} \bibnamefont{Harris}},
     \textit{\bibinfo{bla bla bla}{Conduction threshold and pinning frequency of magnetically induced Wigner solid}},
   \bibinfo{journal}{Phys. Rev. Lett.}
  \textbf{\bibinfo{volume}{66}}, \bibinfo{pages}{3285-3288} (\bibinfo{year}{1991}).

\bibitem[{\citenamefont{Chen et~al.}(2006)\citenamefont{Chen, Sambandamurthy,
  Wang, Lewis, Engel, Tsui, Ye, Pfeiffer, and West}}]{ChenMeltingsolid06}
\bibinfo{author}{\bibfnamefont{Y.~P.} \bibnamefont{Chen}},
  \bibinfo{author}{\bibfnamefont{G.}~\bibnamefont{Sambandamurthy}},
  \bibinfo{author}{\bibfnamefont{Z.~H.} \bibnamefont{Wang}},
  \bibinfo{author}{\bibfnamefont{R.~M.} \bibnamefont{Lewis}},
  \bibinfo{author}{\bibfnamefont{L.~W.} \bibnamefont{Engel}},
  \bibinfo{author}{\bibfnamefont{D.~C.} \bibnamefont{Tsui}},
  \bibinfo{author}{\bibfnamefont{P.~D.} \bibnamefont{Ye}},
  \bibinfo{author}{\bibfnamefont{L.~N.} \bibnamefont{Pfeiffer}},
  \bibnamefont{and} \bibinfo{author}{\bibfnamefont{K.~W.} \bibnamefont{West}},
     \textit{\bibinfo{bla bla bla}{Melting of a 2D quantum electron solid in high magnetic field }},
  \bibinfo{journal}{Nature Phys.} \textbf{\bibinfo{volume}{2}},
  \bibinfo{pages}{452-455} (\bibinfo{year}{2006}).

\bibitem[{ele()}]{electronsolidreview}
\bibinfo{note}{See for a review M. Shayegan, in Perspectives in Quantum Hall
  Effects, edited by S. Das Sarma and A. Pinczuk (Wiley, New York, 1997), Chap.
  9}.

\bibitem[{\citenamefont{Grimes and
  Adams}(1979)}]{GrimesclassicalsolidPRL.42.795}
\bibinfo{author}{\bibfnamefont{C.~C.} \bibnamefont{Grimes}} \bibnamefont{and}
  \bibinfo{author}{\bibfnamefont{G.}~\bibnamefont{Adams}},
     \textit{\bibinfo{bla bla bla}{Evidence for a liquid-to-crystal phase transition in a classical, two-dimensional sheet of electrons}},
  \bibinfo{journal}{Phys. Rev. Lett.} \textbf{\bibinfo{volume}{42}},
  \bibinfo{pages}{795-798} (\bibinfo{year}{1979}).

\bibitem[{\citenamefont{Bockrath}(2008)}]{Bockrath.2008}
\bibinfo{author}{\bibfnamefont{V.~V.} \bibnamefont{Deshpande}} \bibnamefont{and}
  \bibinfo{author}{\bibfnamefont{M}~\bibnamefont{Bockrath}},
     \textit{\bibinfo{bla bla bla}{The one-dimensional Wigner crystal in carbon nanotubes}},
  \bibinfo{journal}{Nature Phys.} \textbf{\bibinfo{volume}{4}},
  \bibinfo{pages}{314-318} (\bibinfo{year}{2008}).


\bibitem[{\citenamefont{Yu and Yang}(2002)}]{Yu02}
\bibinfo{author}{\bibfnamefont{Y.}~\bibnamefont{Yu}} \bibnamefont{and}
  \bibinfo{author}{\bibfnamefont{S.}~\bibnamefont{Yang}},
     \textit{\bibinfo{bla bla bla}{Effect of the tilted field in fractional quantum Hall systems: numerical studies for the solid-liquid transition}},
  \bibinfo{journal}{Phys. Rev. B} \textbf{\bibinfo{volume}{66}},
  \bibinfo{pages}{245318} (\bibinfo{year}{2002}).

\bibitem[{\citenamefont{Pan et~al.}(2005)\citenamefont{Pan, Cs\'{a}thy, Tsui,
  Pfeiffer, and West}}]{pan05tiltholesolid}
\bibinfo{author}{\bibfnamefont{W.}~\bibnamefont{Pan}},
  \bibinfo{author}{\bibfnamefont{G.~A.} \bibnamefont{Cs\'{a}thy}},
  \bibinfo{author}{\bibfnamefont{D.~C.} \bibnamefont{Tsui}},
  \bibinfo{author}{\bibfnamefont{L.~N.} \bibnamefont{Pfeiffer}},
  \bibnamefont{and} \bibinfo{author}{\bibfnamefont{K.~W.} \bibnamefont{West}},
     \textit{\bibinfo{bla bla bla}{Transition from a fractional quantum Hall liquid to an electron solid at Landau level filling $\nu=1/3$  in tilted magnetic fields}},
  \bibinfo{journal}{Phys. Rev. B} \textbf{\bibinfo{volume}{71}},
  \bibinfo{eid}{035302} (\bibinfo{year}{2005}).

\bibitem[{\citenamefont{Santos et~al.}(1992)\citenamefont{Santos, Suen,
  Shayegan, Li, Engel, and Tsui}}]{Santos92holeone3}
\bibinfo{author}{\bibfnamefont{M.~B.} \bibnamefont{Santos}},
  \bibinfo{author}{\bibfnamefont{Y.~W.} \bibnamefont{Suen}},
  \bibinfo{author}{\bibfnamefont{M.}~\bibnamefont{Shayegan}},
  \bibinfo{author}{\bibfnamefont{Y.~P.} \bibnamefont{Li}},
  \bibinfo{author}{\bibfnamefont{L.~W.} \bibnamefont{Engel}}, \bibnamefont{and}
  \bibinfo{author}{\bibfnamefont{D.~C.} \bibnamefont{Tsui}},
     \textit{\bibinfo{bla bla bla}{Observation of a reentrant insulating phase near the 1/3 fractional quantum Hall liquid in a two-dimensional hole system}},
  \bibinfo{journal}{Phys. Rev. Lett.} \textbf{\bibinfo{volume}{68}},
  \bibinfo{pages}{1188-1191} (\bibinfo{year}{1992}).

\bibitem[{\citenamefont{Cs\'{a}thy et~al.}(2005)\citenamefont{Cs\'{a}thy, Noh,
  Tsui, Pfeiffer, and West}}]{csathyholephasediagram04}
\bibinfo{author}{\bibfnamefont{G.~A.} \bibnamefont{Cs\'{a}thy}},
  \bibinfo{author}{\bibfnamefont{H.}~\bibnamefont{Noh}},
  \bibinfo{author}{\bibfnamefont{D.~C.} \bibnamefont{Tsui}},
  \bibinfo{author}{\bibfnamefont{L.~N.} \bibnamefont{Pfeiffer}},
  \bibnamefont{and} \bibinfo{author}{\bibfnamefont{K.~W.} \bibnamefont{West}},
     \textit{\bibinfo{bla bla bla}{Magnetic-Field-induced insulating phases at large $r_{s}$}},
  \bibinfo{journal}{Phys. Rev. Lett.} \textbf{\bibinfo{volume}{94}},
  \bibinfo{eid}{226802} (\bibinfo{year}{2005}).

\bibitem[{\citenamefont{Jiang et~al.}(1990)\citenamefont{Jiang, Willett,
  Stormer, Tsui, Pfeiffer, and West}}]{Jiang90}
\bibinfo{author}{\bibfnamefont{H.~W.} \bibnamefont{Jiang}},
  \bibinfo{author}{\bibfnamefont{R.~L.} \bibnamefont{Willett}},
  \bibinfo{author}{\bibfnamefont{H.~L.} \bibnamefont{Stormer}},
  \bibinfo{author}{\bibfnamefont{D.~C.} \bibnamefont{Tsui}},
  \bibinfo{author}{\bibfnamefont{L.~N.} \bibnamefont{Pfeiffer}},
  \bibnamefont{and} \bibinfo{author}{\bibfnamefont{K.~W.} \bibnamefont{West}},
     \textit{\bibinfo{bla bla bla}{Quantum liquid versus electron solid around $\nu =1/5$ Landau-level filling}},
  \bibinfo{journal}{Phys. Rev. Lett.} \textbf{\bibinfo{volume}{65}},
  \bibinfo{pages}{633-636} (\bibinfo{year}{1990}).

\bibitem[{\citenamefont{Willett et~al.}(1988)\citenamefont{Willett, Stormer,
  Tsui, Pfeiffer, West, and Baldwin}}]{Willett88}
\bibinfo{author}{\bibfnamefont{R.~L.} \bibnamefont{Willett}},
  \bibinfo{author}{\bibfnamefont{H.~L.} \bibnamefont{Stormer}},
  \bibinfo{author}{\bibfnamefont{D.~C.} \bibnamefont{Tsui}},
  \bibinfo{author}{\bibfnamefont{L.~N.} \bibnamefont{Pfeiffer}},
  \bibinfo{author}{\bibfnamefont{K.~W.} \bibnamefont{West}}, \bibnamefont{and}
  \bibinfo{author}{\bibfnamefont{K.~W.} \bibnamefont{Baldwin}},
     \textit{\bibinfo{bla bla bla}{Termination of the series of fractional quantum hall states at small filling factors}},
  \bibinfo{journal}{Phys. Rev. B} \textbf{\bibinfo{volume}{38}},
  \bibinfo{pages}{7881-7884} (\bibinfo{year}{1988}).

    \bibitem[{\citenamefont{D. Shahar, D. C. Tsui, M. Shayegan, R. N. Bhatt, and J. E. Cunningham}(1989)}]{Shahar}
\bibinfo{author}{\bibfnamefont{D.} \bibnamefont{Shahar}},
  \bibinfo{author}{\bibfnamefont{D.C.} \bibnamefont{Tsui}},
  \bibinfo{author}{\bibfnamefont{M.} \bibnamefont{Shayegan}},
    \bibinfo{author}{\bibfnamefont{R.N.} \bibnamefont{Bhatt}},
 \bibnamefont{and}
  \bibinfo{author}{\bibfnamefont{J.E.}~\bibnamefont{Cunningham}},
     \textit{\bibinfo{bla bla bla}{Universal conductivity at the quantum Hall liquid to insulator transition}},
  \bibinfo{journal}{Phys. Rev. Lett.} \textbf{\bibinfo{volume}{74}},
  \bibinfo{pages}{4511-4514} (\bibinfo{year}{1995}).

\bibitem[{\citenamefont{Leadley et~al.}(1997)\citenamefont{Leadley, Nicholas,
  Maude, Utjuzh, Portal, Harris, and Foxon}}]{LeadleyCompskyrmion97}
\bibinfo{author}{\bibfnamefont{D.~R.} \bibnamefont{Leadley}},
  \bibinfo{author}{\bibfnamefont{R.~J.} \bibnamefont{Nicholas}},
  \bibinfo{author}{\bibfnamefont{D.~K.} \bibnamefont{Maude}},
  \bibinfo{author}{\bibfnamefont{A.~N.} \bibnamefont{Utjuzh}},
  \bibinfo{author}{\bibfnamefont{J.~C.} \bibnamefont{Portal}},
  \bibinfo{author}{\bibfnamefont{J.~J.} \bibnamefont{Harris}},
  \bibnamefont{and} \bibinfo{author}{\bibfnamefont{C.~T.} \bibnamefont{Foxon}},
     \textit{\bibinfo{bla bla bla}{Fractional quantum Hall effect measurements at zero g factor}},
  \bibinfo{journal}{Phys. Rev. Lett.} \textbf{\bibinfo{volume}{79}},
  \bibinfo{pages}{4246-4249} (\bibinfo{year}{1997}).

  \bibitem[{\citenamefont{Kang}(1997)\citenamefont{W. Kang, J. B. Young, S. T. Hannahs, E. Palm, K. L. Campman, and A. C. Gossard}}]{Kang97}
\bibinfo{author}{\bibfnamefont{W.} \bibnamefont{Kang}},
  \bibinfo{author}{\bibfnamefont{J.B.} \bibnamefont{Young}},
  \bibinfo{author}{\bibfnamefont{S.T.} \bibnamefont{Hannahs}},
  \bibinfo{author}{\bibfnamefont{E.} \bibnamefont{Palm}},
  \bibinfo{author}{\bibfnamefont{K.L.} \bibnamefont{Campman}},
  \bibnamefont{and} \bibinfo{author}{\bibfnamefont{A.C.} \bibnamefont{Gossard}},
     \textit{\bibinfo{bla bla bla}{Evidence for a spin transition in the v=2/5 fractional quantum Hall effect }},
  \bibinfo{journal}{Phys. Rev. B.} \textbf{\bibinfo{volume}{56}},
  \bibinfo{pages}{12776-12779} (\bibinfo{year}{1997}).


\bibitem[{\citenamefont{DasSarma et~al.}(2000)\citenamefont{}}]{DasSarmaHwang2000}
\bibinfo{author}{\bibfnamefont{S.}~\bibnamefont{Das Sarma}}
 \bibnamefont{and} \bibinfo{author}{\bibfnamefont{E.~H.} \bibnamefont{Hwang}},
    \textit{\bibinfo{bla bla bla}{Parallel magnetic field induced giant magnetoresistance in low density quasi-two-dimensional layers}},
  \bibinfo{journal}{Phys. Rev. Lett.} \textbf{\bibinfo{volume}{84}},
  \bibinfo{pages}{5596-5599} (\bibinfo{year}{2000}).

\bibitem[{\citenamefont{Usher et~al.}(1990)\citenamefont{Usher, Nicholas,
  Harris, and Foxon}}]{Usher91}
\bibinfo{author}{\bibfnamefont{A.}~\bibnamefont{Usher}},
  \bibinfo{author}{\bibfnamefont{R.~J.} \bibnamefont{Nicholas}},
  \bibinfo{author}{\bibfnamefont{J.~J.} \bibnamefont{Harris}},
  \bibnamefont{and} \bibinfo{author}{\bibfnamefont{C.~T.} \bibnamefont{Foxon}},
     \textit{\bibinfo{bla bla bla}{Observation of magnetic excitons and spin waves in activation studies of a two-dimensional electron gas}},
  \bibinfo{journal}{Phys. Rev. B} \textbf{\bibinfo{volume}{41}},
  \bibinfo{pages}{1129-1134} (\bibinfo{year}{1990}).

\bibitem[{\citenamefont{Stopa and Das~Sarma}(1992)}]{StopasquareQW92}
\bibinfo{author}{\bibfnamefont{M.~P.} \bibnamefont{Stopa}} \bibnamefont{and}
  \bibinfo{author}{\bibfnamefont{S.}~\bibnamefont{Das~Sarma}},
     \textit{\bibinfo{bla bla bla}{Density scaling and optical properties of semiconductor parabolic and square quantum wells}},
  \bibinfo{journal}{Phys. Rev. B} \textbf{\bibinfo{volume}{45}},
  \bibinfo{pages}{8526-8534} (\bibinfo{year}{1992}).

\bibitem[{\citenamefont{Stern and Schulman}(1985)}]{Stern85}
\bibinfo{author}{\bibfnamefont{F.}~\bibnamefont{Stern}} \bibnamefont{and}
  \bibinfo{author}{\bibfnamefont{J.~N.} \bibnamefont{Schulman}},
  \bibinfo{journal}{Superlattices and Microstructures}
     \textit{\bibinfo{bla bla bla}{Calculated effects of interface grading in GaAs-Ga$_{1-X}$Al$_{X}$As quantum wells }},
  \textbf{\bibinfo{volume}{1}}, \bibinfo{pages}{303-305} (\bibinfo{year}{1985}).

\bibitem[{\citenamefont{Stopa and
  Das~Sarma}(1989)}]{StopaBparaLandaulevelPRB.40.10048}
\bibinfo{author}{\bibfnamefont{M.~P.} \bibnamefont{Stopa}} \bibnamefont{and}
  \bibinfo{author}{\bibfnamefont{S.}~\bibnamefont{Das~Sarma}},
     \textit{\bibinfo{bla bla bla}{Parabolic-quantum-well self-consistent electronic structure in a longitudinal magnetic field: subband depopulation}},
  \bibinfo{journal}{Phys. Rev. B} \textbf{\bibinfo{volume}{40}},
  \bibinfo{pages}{10048-10051} (\bibinfo{year}{1989}).


\end{thebibliography}
\end{document}